\documentclass[showpacs,a4paper,twocolumn]{revtex4}
\usepackage{graphicx}
\begin{document}
\title{Optimal quantum repeaters with doubly entangled state purification}
\author{Chuan Wang$^{1,2}$, Rong-Zhen Jiao$^{1,2}$, Hai-Qiang Ma$^{1,2}$, Yong Zhang$^{1,2}$ and Li Yu$^{1,2}$}
\affiliation{$1$ School of Science, Beijing University of Posts and
Telecommunications, Beijing, 100876, People's Republic of China\\
$2$ Key Laboratory of Optical Communication and Lightwave
Technologies, Beijing University of Posts and Telecommunications,
Beijing, 100876, People's Republic of China}
\date{\today }

\begin{abstract}
Recently, Briegel et al. present a quantum repeaters protocol using
nest entanglement purification for long distance quantum
communication(Physical Review Letters \textbf{81},5932). In this
paper we present a modified scheme for constructing an optimal
quantum repeater using doubly entangled photon pairs that overcomes
the limitations of quantum communication via noisy channels. Based
on the imperfect quantum operations and classical communications,
the polarization entangled quantum channels can be built with a
better efficiency.

\end{abstract}

 \maketitle

\section{Introduction}\label{ss1}

The idea of quantum key distribution(QKD) was first proposed by
Bennett and Brassard in the year 1984 \cite{BB84}. Later, it is
proved to be unconditional secure \cite{Lo,Shor}. Two fundamental
principles, the uncertain principle and non-clone theorem, protect
the quantum signal not to be eavesdropped without been discovered by
the authorized parties. QKD also provides such a secure way for
creating private keys with which Alice and Bob can exchange the
secret message securely using the proven secure crypto-system and it
has progressed quickly in the following decades
\cite{ekert91,b92,bbm92,bw,entangle,long,bid}. The main obstacle of
long distance quantum communication is the loss of quantum signals
in noisy channels between the communication parties. After long time
research of experimental quantum communication, the length of
quantum communication using single photons is bounded at about 150km
both in free space \cite{Ursin} and in optical fibers \cite{Guo}.
The number of trials of quantum signals is exponentially enlarged as
the length of the channel increased. Another problem of long
distance quantum communication is the increasing of bit error rate
in accordance with the enlarging of the channel distance. Thus the
fidelity of the quantum state is limited in quantum communication.

In 1998, Briegel et al. proposed the idea of quantum repeaters as
the role of imperfect local operations in quantum communication
which overcomes the error problems and the communication distance
problems in quantum communication via noisy channels
\cite{Briegel,Dur}. The protocol utilizes entangled purification
protocol that was proposed by Bennett in which the Werner states can
be purified They introduce the imperfect local operations to build
long distance quantum channel by connecting the distant nodes. An
entangled photon channel is established by such nested purification
protocols. Then the long distance quantum channel can be built with
a high fidelity.

Later, Duan et al. proposed the theoretical protocol of long
distance QKD using atomic ensembles as quantum repeaters
\cite{Duan}. Recently, remarkable progress has been reported on
quantum repeaters \cite{Zhao,Loock,Yuan}. Meanwhile, Collins et al.
proposed the quantum relay protocols to overcome the problems in
long distance quantum communications \cite{Gisin} and the related
topics also progressed quickly \cite{Jacobs,Gisin2}.

In this study, we first introduce the idea of polarization and
frequency doubly entangled photon states(DEPs) to realize the photon
transmission and entanglement purification process and then present
an optimal quantum repeater protocol that allows the creation of
quantum channels using DEPs over arbitrary long distance. The paper
is organized as follows: in section \ref{ss2}, we introduce the DEPs
purification protocol to construct the optimal quantum repeater
protocol by imperfect local operations. In section \ref{ss3}, we
analyze the efficiency and the main features of the optimal quantum
repeaters protocol utilize the generic error model. The last section
is our conclusion and acknowledgement.

\section{Quantum repeaters with DEPs purification protocol}\label{ss2}

Entanglement purification protocol based on DEPs and its modified
version were proposed by Xiao and Wang et al \cite{Xiao,Wang}. This
protocol consists of two steps: first step is the bit-flip error
correction process and second step is the phase-flip error
correction process.

The doubly entangled photon state was experimentally realized by
Ravaro et al. \cite{Ravaro} and it can be written in the formula:
\begin{equation}
\vert \Phi^{+}_{ab}\rangle
=\frac{1}{\sqrt{2}}(|H,\omega_{s}\rangle|H,\omega_{i}\rangle +
|V,\omega_{s'}\rangle|V,\omega_{i'}\rangle),
\end{equation}
here $|H\rangle$ and $|V\rangle$ represent the horizontal and the
vertical polarizations of the photons respectively, $\omega_{s(s')}$
and $\omega_{i(i')}$ correspond to the frequencies of entangled
photons. The transmission of DEPs through noisy channels in quantum
communication will cause the polarizations of each photon effected
by the channel noise. Then the initial state will change to a Werner
state which is represented as
\begin{eqnarray}
\rho & =&
F|\Phi^{+}_{ab}\rangle\langle\Phi^{+}_{ab}|+\frac{1-F}{7}|\Phi^{-}_{ab}\rangle\langle\Phi^{-}_{ab}|
+\frac{1-F}{7}|\Psi^{\pm}_{ab}\rangle\langle\Psi^{\pm}_{ab}| \nonumber \\
& + &
\frac{1-F}{7}|\Gamma^{\pm}_{ab}\rangle\langle\Gamma^{\pm}_{ab}|
+\frac{1-F}{7}|\Upsilon^{\pm}_{ab}\rangle\langle\Upsilon^{\pm}_{ab}|,
\end{eqnarray}\label{e1}
here the coefficient
$F=\langle\Phi^{+}_{ab}|\rho|\Phi^{+}_{ab}\rangle$ is the fidelity
of the state $\rho$ relative to $|\Phi^{+}_{ab}\rangle$ and the
eight states in Equ.(2) are
\begin{eqnarray}
|\Phi^{\pm}_{ab}\rangle=\frac{1}{\sqrt{2}}(|H,\omega_{s}\rangle|H,\omega_{i}\rangle
\pm|V,\omega_{s'}\rangle|V,\omega_{i'}\rangle);\\
|\Psi^{\pm}_{ab}\rangle=\frac{1}{\sqrt{2}}(|H,\omega_{s}\rangle|V,\omega_{i}\rangle
\pm|V,\omega_{s'}\rangle|H,\omega_{i'}\rangle);\\
|\Gamma^{\pm}_{ab}\rangle=\frac{1}{\sqrt{2}}(|V,\omega_{s}\rangle|H,\omega_{i}\rangle
\pm|H,\omega_{s'}\rangle|V,\omega_{i'}\rangle);\\
|\Upsilon^{\pm}_{ab}\rangle=\frac{1}{\sqrt{2}}(|V,\omega_{s}\rangle|V,\omega_{i}\rangle
\pm|H,\omega_{s'}\rangle|H,\omega_{i'}\rangle)
\end{eqnarray}

\begin{figure}
\includegraphics[width=8cm,angle=0]{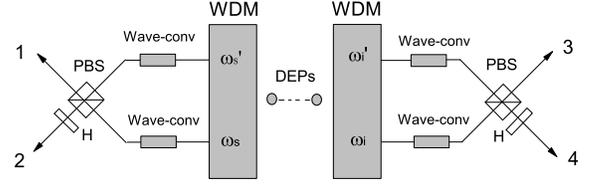}
\caption{ The purification devices. WDMs are wavelength-division
 multiplexing devices. Wave-conv is the wavelength conversion process. PBS and H are the polarizing beam splitter and
 half wave plate, respectively.} \label{f1}
\end{figure}

In the entanglement purification process, the devices shown in
Fig.\ref{f1} is used. In this step, two photons are sent to the left
and right WDMs respectively. The state has no bit-flip errors if the
two photons come out of port 1 and port 3, respectively. When a
photon comes out of the lower spatial mode (port 2 or port 4), a
bit-flip error takes place. Then the two HWPs on port 2 and port 4
is used to correct the bit-flip errors and the DEPs degenerate to
Bell state. The DEPs with phase-flip errors can be further distilled
in the same way as in the existing schemes, as Ref.
\cite{Wang,Pan2}.

After the first step, the fidelity F of the state $|\Phi^{+}\rangle$
increased to $(4F+3)/7$. When the two steps operations finish, the
state becomes to
\begin{equation}
\rho'''=F'|\Phi^{+}_{ab}\rangle\langle\Phi^{+}_{ab}|+(1-F')|\Psi^{+}_{ab}\rangle\langle\Psi^{+}_{ab}|,
\end{equation}
here the fidelity
\begin{equation}
F'=\frac{(4F+3)^2}{32F^2-8F+25}.
\end{equation}
If the initial coefficient $(4F+3)/7$ is larger than $1/2$, which
means that the initial fidelity $F>1/8$, the fidelity $F'$ after
purification is larger than $(4F+3/)7$ and the entanglement
purification succeeds. So we can perform these purification
operations iteratively to increase the fidelity of the ensemble.

\begin{figure}
\includegraphics[width=8cm,angle=0]{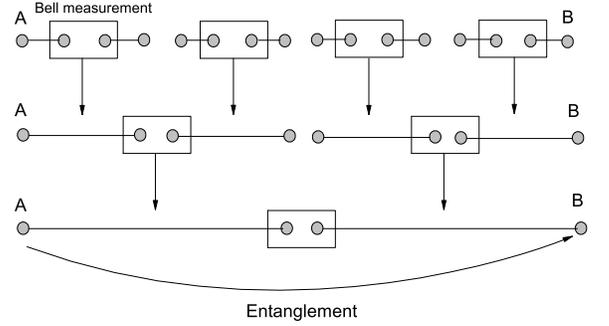}
\caption{ The quantum repeater protocol. Each box represents a node
in the channel, here only three levels $C_{i},C_{2i},C_{3i}$ are
presented.} \label{f2}
\end{figure}

Long distance quantum communication can be realized by quantum
repeaters between distant locations in the communication network. We
have learned the nested purification protocol proposed by Briegel et
al. and improved using our DEPs purification protocols. The scheme
of nest purification quantum repeater is shown in Fig.\ref{f2}. The
two remote parties Alice(A) and Bob(B) need to build an entangled
quantum channel. There are many nodes at distant locations between
them and each node performs an entanglement purification protocol
and a Bell state measurement on their two particles. Then Alice and
Bob could build their quantum channel with a better fidelity.

In our protocol, each neighboring two nodes share a pair of DEPs in
the state $|\Phi^{+}_{ab}\rangle$. The procedure of entanglement
distribution is realized by transferring one particle of DEPs
through the channels. By applying DEPs entanglement purification
protocol on the state between the nodes, the polarization and
frequency doubly entangled photon state degenerate to Bell state.
After that, Alice and $C_{1}$, $C_{1}$ and $C_{2}$, ..., $C_{n-1}$
and Bob are connected by DEPs. Then the channel can be connected by
the method of entanglement swapping. For example, at the first
level, $C_{1}$ apply Bell state measurement on his two particles and
announce the results, as a result, Alice and $C_{2}$ are sharing a
quantum channel by performing a local unitary operation. The same
operations are performed by $C_{i}$ here $i=1,3,5,...$. At the
second level, the $C_{2i}$ nodes perform the same procedures. The
process performs repeatedly by $C_{4i}$, $C_{8i}$ at each level.
Finally, only the two photons at Alice and Bob's side are kept and
the quantum channel is established.

\section{Efficiency of long distance quantum
communication with DEPs quantum repeaters}\label{ss3}

In ideal case, Alice and Bob could establish their quantum channel
with high fidelity by perfect operations and measurements. However,
in realistic conditions, Alice and Bob's imperfect operations and
measurements will reduce the fidelity of the states in the
entanglement purification and Bell state measurement process.

Here in this section we will discuss the efficiency of the quantum
repeaters protocol improved by our DEPs purification protocol. The
imperfect measurement on the single qubit is described as
\begin{eqnarray}
P_{0}=\eta|0\rangle\langle 0|+(1-\eta)|1\rangle\langle 1|.\\
P_{1}=\eta|1\rangle\langle 1|+(1-\eta)|0\rangle\langle 0|.
\end{eqnarray}
Here $|0\rangle$ and $|1\rangle$ represent the quantum state of
single photon respectively and $\eta$ is the parameter of the
projection quality.

In our purification procedures, single qubit operation and state
measurement are needed. The imperfect single qubit operation on the
density matrix of the state is described by the map
\begin{eqnarray}
\rho\longrightarrow \hat{O_{1}}\rho=p_{1}
\hat{O_{1}}^{ideal}\rho+\frac{1-p_{1}}{2}tr_{1}{\rho}\otimes I_{1}.
\end{eqnarray}
Here $O_{1}^{ideal}$ is the ideal operations. $I_{1}$ is the
subspace where the ideal operation acts.

We analyze the two procedures of nest DEPs purification protocol in
the error models shown above. The quantum repeater protocol consists
two element: entanglement purification and the connection of photon
pairs. The protocol takes place where the entanglement channels
between the nodes are in the Werner state:
\begin{eqnarray}
\rho & =&
F|\Phi^{+}_{ab}\rangle\langle\Phi^{+}_{ab}|+\frac{1-F}{7}|\Phi^{-}_{ab}\rangle\langle\Phi^{-}_{ab}|
+\frac{1-F}{7}|\Psi^{\pm}_{ab}\rangle\langle\Psi^{\pm}_{ab}| \nonumber \\
& + &
\frac{1-F}{7}|\Gamma^{\pm}_{ab}\rangle\langle\Gamma^{\pm}_{ab}|
+\frac{1-F}{7}|\Upsilon^{\pm}_{ab}\rangle\langle\Upsilon^{\pm}_{ab}|,
\end{eqnarray}
The fidelity of the state is
$F=\langle\Psi^{+}|\rho|\Psi^{+}\rangle$.

Considering the imperfect operations, the nodes between Alice and
Bob will perform Bell state measurement on their two particles. Then
Alice's and Bob's particles become entangled with a certain
fidelity. Briegel et al. illustrate that the fidelity decrease
exponentially with N. By connecting the N neighboring pairs.

Note that in the optimal quantum purification protocol, the quantum
repeater protocol can be described as follows. In the first step
purification process, two half wave plates are operated on the port
2 and 4 to correct the bit-flip errors. In this process, the density
matrix $\rho$ of the ensemble changes to $\rho'$.
\begin{eqnarray}
\rho' & =&
F|\Phi^{+}_{ab}\rangle\langle\Phi^{+}_{ab}|+\frac{1-F}{7}|\Phi^{-}_{ab}\rangle\langle\Phi^{-}_{ab}|
+p_{1}\frac{1-F}{7}|\Phi^{\pm}_{ab}\rangle\langle\Phi^{\pm}_{ab}| \nonumber \\
& + &
p_{1}\frac{1-F}{7}|\Phi^{\pm}_{ab}\rangle\langle\Phi^{\pm}_{ab}|
+p_{1}^{2}\frac{1-F}{7}|\Phi^{\pm}_{ab}\rangle\langle\Phi^{\pm}_{ab}|+\emph{C}(p_{1}),
\end{eqnarray}
here $\emph{C}(p_{1})$ denotes the errors introduced by the
imperfect single qubit operations. Also the density matrix can be
written as
\begin{eqnarray}
\rho' & =&
(F+2p_{1}\frac{1-F}{7}+p_{1}^{2}\frac{1-F}{7})|\Phi^{+}_{ab}\rangle\langle\Phi^{+}_{ab}|
+(\frac{1-F}{7}\nonumber\\
& + &
2p_{1}\frac{1-F}{7}+p_{1}^{2}\frac{1-F}{7})|\Phi^{-}_{ab}\rangle\langle\Phi^{-}_{ab}|+\emph{C}(p_{1})
\end{eqnarray}

Then we perform the second step purification and correct the
phase-flip errors. Another pair of entangled photon pairs selected
from the ensemble is needed. Then four half-wave plates operate on
each of the photons in this process. The density matrix becomes to
\begin{eqnarray}
\rho'' & =& p_{1}^4 F'|\Phi^{+}_{ab}\rangle\langle\Phi^{+}_{ab}|
+p_{1}^{4} (1-F')
|\Psi^{+}_{ab}\rangle\langle\Psi^{+}_{ab}|+\emph{C}(p_{1})
\end{eqnarray}
where
\begin{eqnarray}
F'=\frac{(F+2p_{1}\frac{1-F}{7}+p_{1}^{2}\frac{1-F}{7})^2}{(F+2p_{1}\frac{1-F}{7}+p_{1}^{2}\frac{1-F}{7})^2
+(\frac{1-F}{7} + 2p_{1}\frac{1-F}{7}+p_{1}^{2}\frac{1-F}{7})^2}
\end{eqnarray}

Using DEPs entanglement purification protocol, $F''$ is solved as
the final fidelity of the new pair after the entanglement
purification process.
\begin{eqnarray}
F''=\frac{p_{1}^{4}F'[\eta^2+(1-\eta)^2]+(1-p_{1}^4)/64}{p_{1}^{4}F'[\eta^2+(1-\eta)^2]
+p_{1}^{4}(1-F')2\eta(1-\eta)+(1-p_{1}^4)/8}
\end{eqnarray}

\begin{figure}
\includegraphics[width=8cm,angle=0]{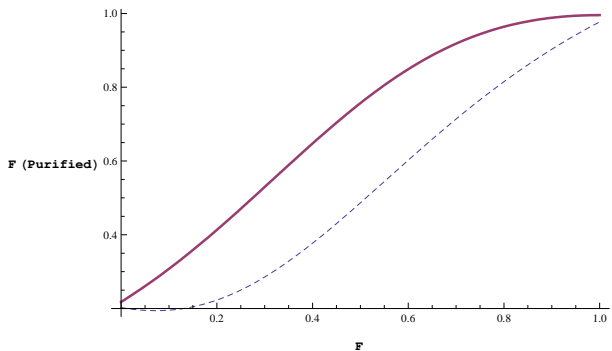}
\caption{ The efficiency of entanglement purification protocols.
Thick line is the DEPs purification efficiency and the dashed line
is Bennett's method for purification at one time in quantum
repeaters protocol. Parameters are $p_{1}=p_{2}=0.99$, $\eta=1$}
\label{f3}
\end{figure}

The efficiency of the purification scheme in quantum repeaters is
shown in Fig.\ref{f3}.
We can see from the curve that DEPs purification is better at each
time purification. So the efficiency of this protocol is higher than
the traditional quantum repeater protocols.


Here we introduce the modified DEPs entanglement purification into
the nest purification quantum repeater protocol. Comparing with the
traditional protocols, our scheme avoids using the two qubit quantum
operations which is difficult to realize in experiment. When Alice
and Bob start the long distance quantum communication, they need to
build a quantum channel. Using the DEPS nest purification protocols,
they first distribute the DEPs between the nodes. Then the nodes
$C_{i}$ perform the entanglement purifications on the photon pairs
at each level and finally they share the state with fidelity $F''$.
The last step is to perform the Bell state measurement and quantum
teleportation on the two particles at the nodes. These steps iterate
at the nodes $C_{2i}$, $C_{4i}$, $C_{8i}$ and so on. At last, a high
quantum channel is established.

\section{Conclusion }\label{ss5}

In this study, we have proposed an optimal quantum repeaters
protocol using DEPs entanglement purification protocol and show that
creating DEPs channels via a noisy channel is possible. The
additional degree of freedom of the DEPs increases the efficiency of
purification, so it makes the two elements in the quantum repeater
improved. Compared with the Bennett purification protocols, the
efficiency of the quantum repeaters using DEPs purification with
imperfect quantum operations are discussed.

\section*{ACKNOWLEDGMENTS}

This work is supported by the National Natural Science Foundation of
China under Grant No. 10704010, 10805006.

\end{document}